\documentclass[10pt]{article}
\usepackage{amsmath,amsfonts,amssymb,amsthm,amsopn,amscd}
\usepackage{bbm}
\usepackage[dvips]{graphicx}
\usepackage{inputenc}
\usepackage{hyperref}
\usepackage{color}
\usepackage{colortbl}
\usepackage{subfigure}

\usepackage{cite}

\usepackage[labelfont=bf,labelsep=period,justification=raggedright]{caption}

\makeatletter
\renewcommand{\@biblabel}[1]{\quad#1.}
\makeatother

\date{}

\definecolor{turquoise}{rgb}{0.00,0.40,0.50}    

\def\corrfirst#1{#1} 	

\def\corr#1{{#1}} 	

\begin{document}

\noindent {\it PLoS One}, in press, 2010.

	\begin{flushleft}
	{\Large
	\textbf{Can power-law scaling and neuronal avalanches arise
	from stochastic dynamics?}
	}
	\\
	Jonathan Touboul $^{1,2,\ast}$, 
	Alain Destexhe$^{3}$, 
	\\
	{\bf 1} Department of Mathematics, University of Pittsburgh, Pittsburgh PA, USA
	\\
	{\bf 2} Laboratory of Mathematical Physics, The Rockefeller University, New York, NY USA
	\\
	{\bf 3} Unit\'e de Neurosciences Int\'egratives et Computationnelles (UNIC), UPR CNRS 2191, 91198 Gif-sur-Yvette, France.
	\\
	$\ast$ E-mail: jonathan.touboul@gmail.com
	\end{flushleft}

\section*{Abstract}

The presence of self-organized criticality in biology is often evidenced \corrfirst{by
a power-law scaling of event size distributions, which can be measured by linear
regression on logarithmic axes}.  We show here that such a procedure does not
necessarily mean that the system exhibits \corrfirst{self-organized criticality}. We
first provide an analysis of multisite local field potential (LFP) recordings of
brain activity and show that event size distributions defined as negative LFP
peaks can be close to power-law distributions. \corrfirst{However, this result} is not
robust to change in detection threshold, or \corrfirst{when tested using more
rigorous} statistical analyses such as the Kolmogorov--\corrfirst{Smirnov} test. 
Similar power-law scaling is observed for surrogate signals, suggesting that
power-law scaling may be a generic property of thresholded stochastic processes. 
We \corrfirst{next} investigate this problem analytically, and show that, indeed,
\corrfirst{stochastic processes can produce spurious power-law scaling} without the
presence of underlying self-organized criticality.  However, this power-law is
only apparent in logarithmic representations, \corrfirst{and} does not \corrfirst{survive}
more \corrfirst{rigorous} analysis such as the Kolmogorov--\corrfirst{Smirnov} test. 
\corrfirst{The same analysis was also performed on an artificial network known to
display self-organized criticality. In this case, both the graphical
representations and the rigorous statistical analysis reveal with no ambiguity
that the avalanche size is distributed as a power-law}.  We conclude that
logarithmic representations can lead to spurious power-law scaling induced by the
stochastic nature of the phenomenon. \corrfirst{This apparent power-law scaling does
not constitute a proof of self-organized criticality, which should be
demonstrated by more stringent statistical tests.}

\section*{Introduction}

\corr{Many natural complex systems, such as earthquakes or sandpile
avalanches, permanently evolve at a phase transition point, a type
of dynamics called self-organized criticality (SOC)
\cite{bak:96,jensen:98}}.  \corr{SOC} \corrfirst{states are potentially
interesting for neural information processing because they
represent a state consisting of ``avalanches'' of recruitment of
units as opposed to oscillations or waves.  One of the signatures
of such critical states is that the size of the avalanches is
distributed as a power law, which is particularly interesting for
the scale invariance it presents\footnote{\corrfirst{More
precisely, if the probability of observing value x for a given
variable is a power-law, $p(x)=a x^{-\alpha}$, then scaling $x$ by
a constant factor yields to a proportional law: $p(c\,x) = a
c^{-\alpha} x^{-\alpha}$.}}.  Another notable property is the
universality of power-laws in physical phenomena such as phase
transitions. In these cases, the exponent is called the critical
exponent.  Diverse systems show the same critical exponent as they
approach criticality, indicating the same fundamental dynamics.}

\corrfirst{In neuroscience, it is of obvious interest to determine
if the recruitment of activity in neural networks occurs in
power-law distributed avalanches.  This would be \corr{evidence
that the brain may function} according to critical states, rather
than the usual wave-type, oscillatory or stochastic dynamics. 
Moreover, power-law relations are \corr{often associated with}
long-lasting correlations in the system, \corr{as with} the
behavior near critical points.  Indeed, the presence of
self-organized criticality was \corr{inferred} for several
biological systems, including spontaneous brain activity {\it in
vitro}~\cite{beggs-plenz:04} which displays spontaneous bursts of
activity -- or ``neuronal avalanches'' -- separated by silences
(see also \cite{hennig:09} for spontaneous activity in the
retina)}.  The distribution of such events was \corr{identified to
scale} as a power law, which \corr{was taken as} evidence for
self-organized criticality in this system (see also review
by~\cite{jensen:98}).  

To investigate if criticality is important for brain function, the same type of
analysis was also investigated {\it in vivo}, and in particular in awake animals.
However, the difficulty with such analyses is that the activity in awake animals
is much more intense compared to {\it in vitro}~\cite{steriade:01}, with often no
visible ``pause'' in the firing activity, which complicates the definition of
avalanches.  In a first study on awake cats~\cite{bedard-kroger-etal:06}, it was
shown that \corrfirst{although macroscopic variables such as the extracellular local
field potential (LFP) show $1/f$ scaling in power spectra, the underlying
neuronal activity does not show signs of criticality}.  In a second, more recent
study on awake monkeys~\cite{peterman:09}, power-law scaling was apparent from
LFPs when considering the statistics of negative peaks, which are known to be
related to neuronal firing.  This scale-invariant behavior was taken as evidence
for self-organized criticality.  

In the present paper, we attempt to resolve these contradictory observations by
first performing the same analysis on negative LFP peaks in cats, and \corrfirst{using
different statistical tests and models to explain these observations.}  We study
the statistical distribution of avalanche sizes, as well as \corrfirst{the
distribution of the amplitude of} negative peaks in the LFPs (linked to neuronal
firings), positive peaks, and surrogate data.  We then study similar stochastic
problems, and \corrfirst{investigate whether} the results obtained by the experimental
data analysis can also be observed in purely stochastic systems without the
presence of underlying self-organized criticality.  \corrfirst{Eventually, we compare the
results obtained to the analysis of avalanche data produced by a neural network
known to present self-organized criticality \cite{levina-etal:07,levina-etal:09}.}

\section*{Material and Methods}

\subsection*{Experimental Data}

\begin{figure}[!ht]
\begin{center}
 \includegraphics[width=\textwidth]{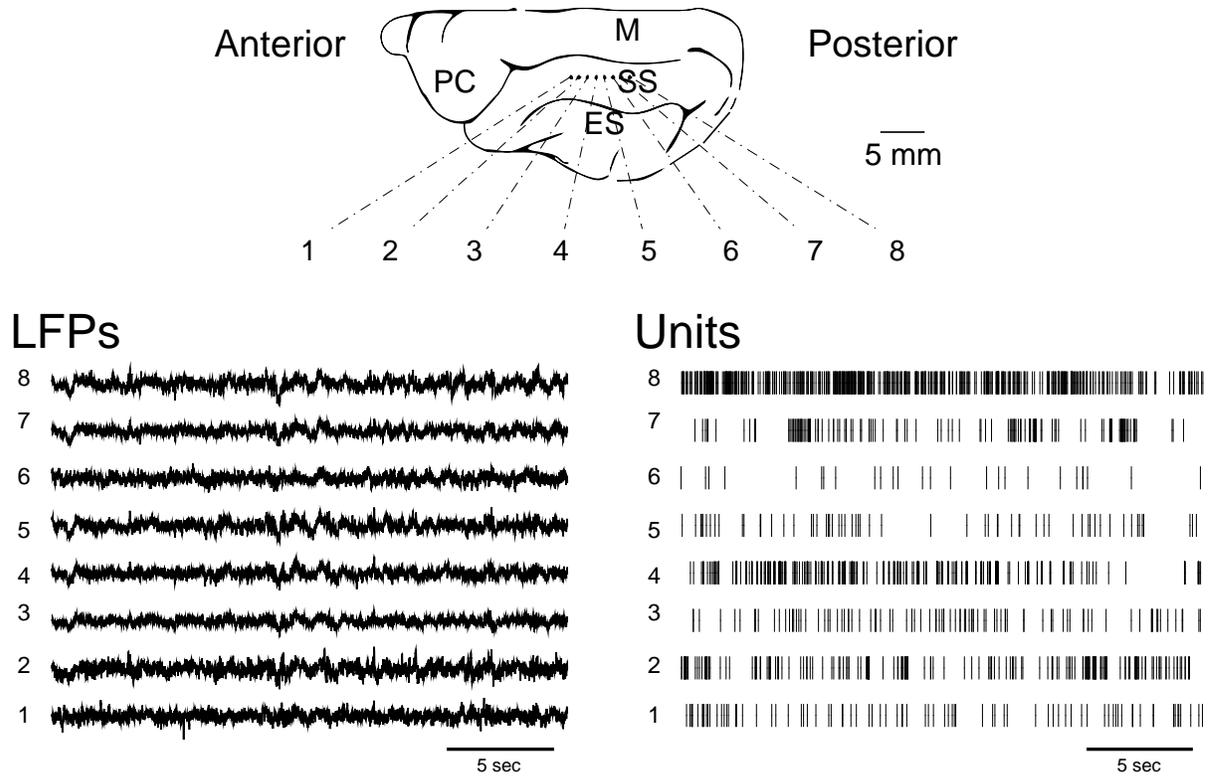}
\end{center}
\caption{{\bf Simultaneous multisite LFP and unit recordings in awake
cats.}  Eight pairs of tungsten electrodes \corrfirst{ (placement
illustrated on top)} were inserted in cat
cerebral cortex (area 5-7, parietal) as described in detail
in \cite{destexhe-contreras-etal:99}.  The system 
simultaneously recorded LFPs (left) and multi-unit activity (right) 
at each pair of electrode.}
\label{exp}
\end{figure}

The experimental data used in the analysis consist of simultaneous
recordings of multisite local field potentials (LFPs) and unit
activity in the parietal cortex of awake cats (see Fig.~\ref{exp}),
which were obtained from a previous
study~\cite{destexhe-contreras-etal:99}.  \corrfirst{A linear array
of 8 bipolar electrodes was chronically implanted in the gray
matter of area 5-7 of cat cerebral cortex, and the state of the
animal was monitored to insure that all recordings were made in
awake conditions \corr{(quiet wakefulness with eyes-open)}. 
Signals were recorded on an eight-channel digital recorder
(Instrutech, Mineola, New York) with internal sampling rate of
11.8~kHz per channel, and 4-pole Bessel filters.  For LFPs, data
were digitized off-line at 250~Hz using the Igor software package
(Wavemetrics, Oregon; A/D board from GW Instruments, Massachusetts;
low pass filter of 100~Hz).  Units were digitized off-line at
10~kHz, and spike sorting and discrimination was performed with the
DataWave software package (DataWave Technologies, Colorado; filters
were 300~Hz high-pass and 5~kHz low-pass).  The data was
transferred to LINUX workstations for analysis.}

\corrfirst{\subsection*{LFP analysis}}

\corrfirst{\subsubsection*{Peak detection}}

\corrfirst{Negative or positive peaks were detected from the LFPs
as follows.  Signals were mean-subtracted and divided by their
standard deviations to obtain comparable amplitude statistics.  To
avoid artifactual peak detection because of occasional slow
components or drifts, the signals were digitally filtered below
15~Hz (high-pass), and the peaks were detected using an adjustable
fixed threshold.  The peak was defined as the extremum of the
ensemble of data points that exceeded the threshold. The detected
peaks were then repositioned in the intact original signal
\corr{(see Fig.~\ref{nLFP})}.  The same method was also used for
detecting positive peaks.}

\corrfirst{\subsubsection*{Avalanche analysis}}

\corrfirst{Avalanches were defined by binning the raster of negative peaks of the LFP
(nLFPs) into time bins of size $\Delta t$ (varied between 4 and 16 ms), and by
defining avalanches as clusters of activity among electrodes, separated by silent
periods (time bins with no activity), in accordance with previous
studies~\cite{beggs-plenz:04,peterman:09}.  The ``size'' of each avalanche was
defined as the sum of the amplitudes of all LFP peaks in the avalanche.  Similar
results were obtained when avalanche size was defined as the total number of
peaks within each avalanche (not shown).}

\corrfirst{\subsubsection*{Surrogate signals}}

\corrfirst{Surrogate signals were generated from the nLFP data sets by shuffling the
occurrence times of the different peaks, while keeping the same distribution of
peak amplitudes.  The occurrence times were replaced by random numbers taken from
a flat distribution.  The avalanche analysis was then performed on this shuffled
data set.  Note that, because shuffling changed the timing of the peaks, the
whole set of avalanches changed.}

\

\corrfirst{\subsection*{Artificial Data}}

\corrfirst{The results of neuronal avalanche analysis recorded in the cat cerebral
cortex will be compared to two types of artificial data sets. From the nature of
the LFPs and the links between unit firing and LFP peaks above a certain
threshold (see the Results section), we will compare the results of the avalanche
analysis of cortical data with two simple stochastic processes (not at
criticality) in order to see if the results observed in the avalanche analysis of
cortical data can be linked with the stochastic nature of the LFPs. We will also
compare the results of the avalanche analysis on experimental data to the
avalanche analysis of a network that presents self-organized criticality. }

\corrfirst{\subsubsection*{Stochastic models}}

\corrfirst{The stochastic processes studied are based on the following
two simple models.}

\paragraph{\corrfirst{The shot noise model}}
\corrfirst{The first stochastic model considered is a high-frequency shot-noise
process consisting of exponential events convolved with a Poisson
process. This process, $V_t$, satisfies the equation}
\begin{equation}\label{eq:ShotNoise}
\tau_m dV_t = -V_t\,dt + \sum_{i=1}^{P} q_i \, dN^{(i)}_t 
\end{equation}
where $\tau_m $ is the characteristic decay time constant of each 
exponential event, $q_i$ is the jump amplitude of each event, and 
$N^{(i)}_t$ are independent Poisson processes.  The solution of
Eq.~\eqref{eq:ShotNoise} can be written as:
\begin{equation}\label{eq:SolShotNoise}
V_t = V_0 \exp\left(-\frac{t}{\tau_m}\right) + \sum_{i=1}^P \sum_{t^{i} \text{times of } N^{(i)}}
\exp\left(-\frac{t-t_i}{\tau_m}\right) ~ .
\end{equation}
Here, the stochastic variable $V_t$ represents the LFP as the 
summation of a large number of randomly-occurring synaptic events, 
each described by a decaying exponential.  

\paragraph{\corrfirst{The Ornstein-Uhlenbeck model}}
In the limit of a high number of Poisson processes with summable
intensities (or in the limit of a finite number of Poisson process with high 
firing rate and suitable scaling on the jump amplitude), the solution of 
equation \eqref{eq:ShotNoise} converges in law towards the solution of the 
equation:
\begin{equation}\label{eq:DiffApprox}
\tau_m dV_t = (\mu-V_t)\,dt + \sigma \,dW_t
\end{equation}
\corrfirst{where $W_t$ is a Wiener process,} $\mu$ is related to the \corrfirst{variables}
$q_i$ and \corrfirst{to} the rates of the Poisson processes. This convergence can be
proved using for instance Donsker's theorem (see e.g. \cite{billingsley:99,
touboul-faugeras:07b}) and is generally called \emph{diffusion approximation}.
The process solution of equation \eqref{eq:DiffApprox} is an Ornstein--Uhlenbeck
process, \corrfirst{given by}:
\begin{equation}\label{eq:SolOU}
V_t = V_0 e^{-t/\tau_m} + \mu (1-e^{-t/\tau_m}) + \frac{\sigma}{\tau_m} \int_{0}^{t} e^{(s-t)/\tau_m}\,dW_s
\end{equation}

\corrfirst{\subsubsection*{Self-organized critical neural network}}

\corrfirst{We finally performed the statistical avalanche size analysis in a situation
where self-organized criticality was known to be present.  We used a model
proposed by Levina and colleagues, which consists of a network of spiking neurons
with dynamical synapses, in which the neuronal avalanches are characterized by a
typical and robust self-organized critical behavior
\cite{levina-etal:07,levina-etal:09}.  The network is composed of $N$ so-called
\emph{perfect integrate-and-fire neurons} that integrate random inputs without
linear effects such as the cell membrane's leak and without nonlinear effects due
to the channels dynamics, and that fire a spike when the membrane potential
reaches a fixed threshold. The spike is transmitted with a fixed delay to all
postsynaptic neurons with a connectivity weight that varies according to the
available reserve of neurotransmitter. This type of network with such dynamic
synapses self-tunes to criticality\cite{levina-etal:07}.}

\

\subsection*{Identifying tail distributions}\label{sec:stats}

\subsubsection*{Power-law and exponential distributions}

Mathematically,  a \corrfirst{continuous} random variable $X$ is said to present a
power-law distribution  if it is drawn from a probability distribution with
density:
\begin{equation}\label{eq:powerlawdensity}
\mathbb{P}(x\leq X\leq x+dx) = a\,x^{-\alpha}  \,dx
\end{equation}
where $\alpha$ is a constant parameter of the distribution known as the
\emph{exponent} or \emph{scaling} parameter, \corrfirst{and $a$ is a normalization
parameter. A discrete power-law random variable has a similar, discretized
version of the probability, that can be written $P(X=k) = a k^{-\alpha}$.} In
practice, few empirical phenomena obey power laws for all values of $X$, and in
general power laws characterize the tail of the distribution, i.e. the
probability distribution of values of $X$ greater than some value
$\corrfirst{x_{\min}}$ . In such cases, we say that the tail of the distribution
follows a power law. \corrfirst{Moreover, the data often show a
truncated power law distribution, i.e. power-law behavior only
over a limited range, $x_{\min}\leq x \leq x_{\max}$.}

In this paper, we are interested in discriminating power-laws \corrfirst{from} another type of
\corrfirst{distribution:} the exponentially-tailed distribution. Random variables with such distributions are
characterized for $x\geq \corrfirst{x_{\min}}$ by an exponential probability density,
\corrfirst{that in the continuous case is given by}:
\begin{equation}\label{eq:explawdensity}
\mathbb{P}(x\leq X\leq x+dx) \corrfirst{=} C e^{-\lambda x}  \,dx
\end{equation}
\corrfirst{where $\lambda$ is the parameter of the exponential law and $C$ is a scaling
parameter.  The discrete law can be written in a similar fashion $P(X=k)=C
\lambda^{-k}$.} Given some experimental data, the problem is to identify the
parameters of the power-law or exponential law \corrfirst{that best fits}, which means
estimating the parameter $\hat{x}_{\min}$, and the power-law exponent
$\hat{\alpha}$ or the exponential-law intensity $\hat{\lambda}$.

\subsubsection*{Parameter evaluations}

Taking the logarithm of the probability density of a power-law random variable,
we obtain $\log(p(x)) = -\alpha \log(x) + \log(a)$. The histogram of the
power-law therefore presents an affine relation in a log-log plot. Similarly, the
exponential distribution's histogram is characterized by an affine relation in a
log-linear plot. For this reason, power-laws in empirical data are often studied
\corrfirst{by} plotting the logarithm\footnote{\corrfirst{In this paper, logarithm (and the
$\log$ function) corresponds to the natural (neperian) logarithm function}} of
the histogram as a function of the logarithm of the values of the random
variable, and doing a linear regression \corrfirst{to fit an affine line to through
the data points (usually using a least-squares algorithm)}.  This \corrfirst{method}
dates back to Pareto in the 19th century (see e.g.  \cite{arnold:83b}).  The
evaluated point $\hat{x}_{\min}$ corresponding to the point where the data start
having a power-law distribution is mostly evaluated visually, but this method is
very sensitive to noise, \corrfirst{and is highly} subjective (see e.g. 
\cite{stoev:06} and references herein).  This widely used technique \corrfirst{(and
similar variations)} generate systematic errors under relatively common
conditions (see e.g.  \cite{clauset-etal:09}).  Moreover, there is not any
evaluation of the goodness of fit obtained \corrfirst{under} the power-law assumption.
In this paper, we prefer to use a maximum likelihood estimator, which is
considered the most reliable of usual estimators (see \cite{clauset-etal:09} for
a comparison of different estimators). It is known to provide an accurate
parameter estimate in the limit of large sample size (see
\cite{barndorff:95,muniruzzaman:57}).

Assume that $\corrfirst{x_{\min}}$, the starting value above which the tail of the
distribution, is known, expressions \corrfirst{giving} the maximum likelihood
estimator and maximal likelihood are well known.  \corrfirst{For the continuous
power-law distribution, the maximum likelihood estimator of the exponent
parameter $\alpha$ corresponding to n data points $x_{i}\geq \corrfirst{x_{\min}}$
is:} 
\[\hat{\alpha} = 1+n \big( \sum_{i=1}^{n}
\log \frac{x_{i}}{\corrfirst{x_{\min}}}\big)^{-1}.\] 
\corrfirst{For the continuous
exponential distribution, the maximum likelihood estimator of the parameter 
$\lambda$ is:}
\corrfirst{ \[\hat{\lambda} =(\langle x \rangle -
\corrfirst{x_{\min}})^{-1},\] 
where $\langle x \rangle = \frac 1 n \sum_{i=1}^n x_i$
is the average value of the observations $x_i$.}

\corrfirst{For the continuous power-law distribution the log-likelihood of the data 
for the estimated parameter value is:} 
\[L(\hat{\alpha} \vert
X)  = n\, \log \left(\frac{\hat{\alpha}-1}{\corrfirst{x_{\min}}}\right) - \hat{\alpha}
\sum_{i=1}^{n}\log\left ( \frac{x_{i}}{\corrfirst{x_{\min}}} \right)\] 
and \corrfirst{for the continuous exponential law}: \[L(\hat{\lambda} \vert X) = n
\log(\hat{\lambda})-\hat{\lambda} \sum_{i=1}^{n} (x_{i}-\corrfirst{x_{\min}}).\]

\corrfirst{For the discrete exponential distribution, the maximum likelihood estimator
has exactly the same expression as that for the continuous exponential law. The
exponent estimator for the discrete power-law (truncated or otherwise) has a more
complex form than that for the continuous power-law, and cannot easily be
expressed as a function of the data (see e.g. \cite{bauke:07}).  The log
likelihood of a sample $(x_{i}; i=1,\ldots,n) \in \mathbbm{N}^n$ is: 
\[L(\alpha) = -\alpha \sum_{i=1}^n \log(x_i)-n \,
\log(\sum_{k=k_{\min}}^{k_{\max}}k^{-\alpha}),\] and the estimated value
$\hat{\alpha}$ is given by the unique value of
$\alpha$ that maximizes the above likelihood function. }

\corrfirst{Therefore, given the samples $(x_i)$ and the value of $x_{\min}$ (and
possibly $x_{\max}$), we have expressions for the estimated power-law or
exponential parameter. The parameter $\hat{x}_{\min}$ is evaluated then by
minimizing the Kolmogorov--\corrfirst{Smirnov} distance}: 
\[KS=\max_{x\geq \corrfirst{x_{\min}}} \vert S(x)-\hat{P}(x) \vert \] 
\corrfirst{where $S(x)$ is the cumulative distribution function CDF of
the data and $\hat{P}(x)$ is the CDF of the theoretical distribution
being fitted for the parameter that best fits the data for $x\geq
\corrfirst{x_{\min}}$), as proposed by Clauset and colleagues in
\cite{clauset-etal:07}.}

\subsubsection*{Goodness-of-fit and p-value validation}

\corrfirst{For a} given data set, we now know how to evaluate the best power-law and
best exponential-law fits. But is \corrfirst{either} fit plausible and accurate? In
order to answer this question, we use \corrfirst{a} standard goodness-of-fit test
which generates a p-value quantifying the \corrfirst{likelihood of obtaining a fit as
good or better than that observed, if the hypothesized distribution is correct}. 
\corrfirst{This method involves sampling the fitted distribution to generate
artificial data sets of size $n$, and then calculating the Kolmogorov--Smirnov
distance between each data-set and the fitted distribution, producing the
distribution of Kolmogorov--Smirnov distances expected if the fitted distribution
is the true distribution of the data.  A p-value is then calculated as the
proportion of artificial data showing a poorer fit than fitting the observed data
set.} When this value is close to $1$, the data set can be considered to be drawn
from the \corrfirst{fitted distribution}, and if not, the hypothesis might be
rejected. The smallest p-values often considered to validate the statistical test
are taken between $0.1$ and $0.01$.  \corrfirst{These values are computed following
the method described in \cite{clauset-etal:09}, which in particular
\corrfirst{involves} generating artificial samples through a Monte-Carlo procedure.}

\subsubsection*{\corrfirst{Direct comparison of models}}

\corrfirst{The methods described above provide the better possible fit for a data set with different probability laws and and the statistical relevance of the model fitted to explain the data set. However, in the case where neither model is rejected by the p-value test, these procedures do not allow to quantify which model provides the better fit.}

\corrfirst{Several methods have been proposed to directly compare
models, such as cross validation \cite{stone:74}, fully bayesian
approaches \cite{kass-raftery:95}, minimum description length
\cite{grunwald:07} and the classical log likelihood
ratio\cite{mood:63,vuong:89}. The latter, our method of choice, is
of particular interest because of the Neyman--Pearson lemma
establishing its optimality in certain
conditions\cite{neyman-pearson:33}. This method compares the
likelihood of the fit for each model, and selects the model with
the greater likelihood.} \corr{ The sign of the log likelihood
ratio gives an indication of the model that best fits the data
(note that its amplitude in absolute value does not have a direct
interpretation), but as other statistical quantities, it is
sensitive to noise. The significance of this test therefore needs
to be evaluated, and depends on the size of the sample and on the
empirical standard deviation of the difference between the log
likelihoods of each model (see \cite{vuong:89}). This significance test
gives a scalar value (also called p-value) between $0$ and $1$. If this value is close to zero, then it is unlikely that the sign of the log likelihood ratio is a result of fluctuations. In that case, it is considered that the sign of the log likelihood ratio is a reliable indicator of which model is the better fit to the data. If it is close to one, the sign is not reliable and the test does not favor either model over the other.}

\corrfirst{Note that this method compares fits on a given same data set, which requires in particular the use of the same $x_{\min}$ in both models. For this test, we fix $x_{\min}$ to the mean of the two $x_{\min}$ estimated for each law, thereby giving an advantage to the model that fits more of the data.}

\section*{Results}

\subsection*{\corrfirst{Avalanche analysis of LFPs from cat cerebral cortex}}

\begin{figure}[ht!]
\begin{center}
 \includegraphics[width=0.8\textwidth]{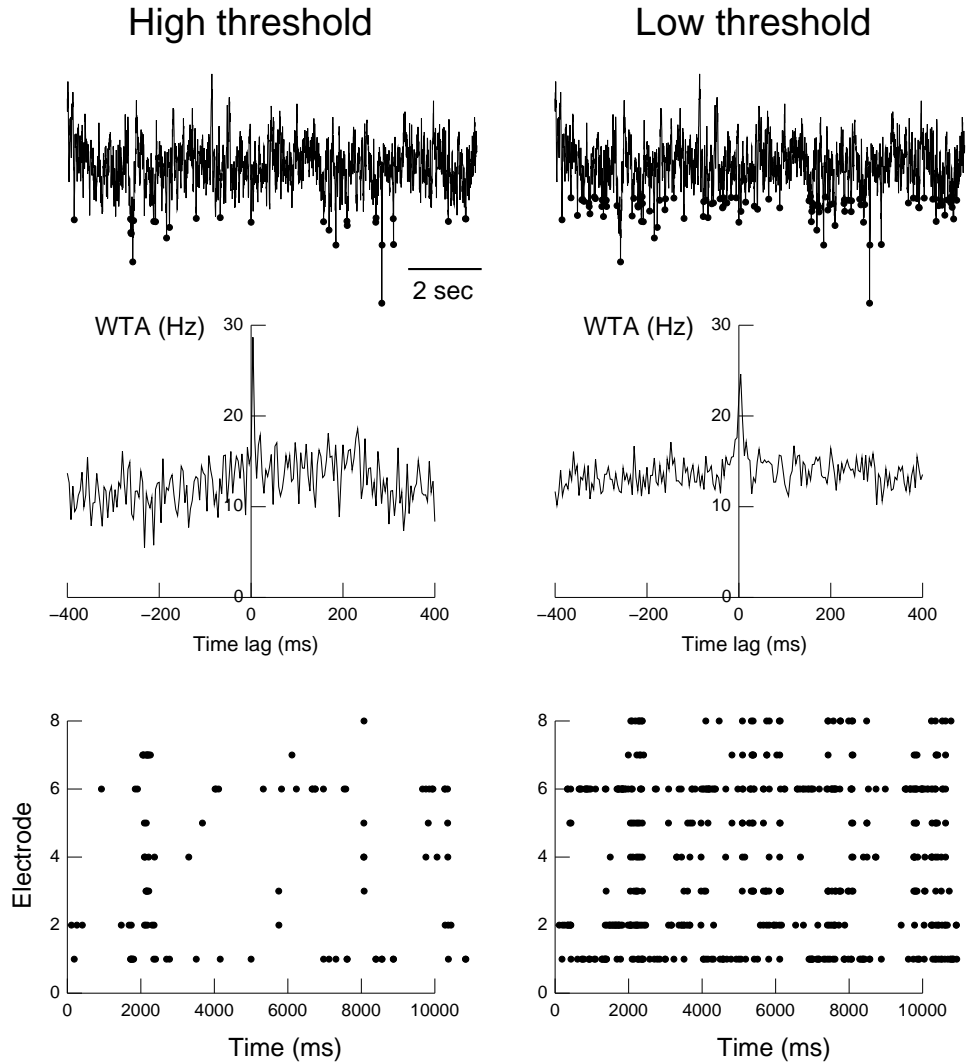}
\end{center}
\caption{{\bf Detection of negative peaks in local field potentials and
their relation with neuronal activity.}  Top: detection of negative
LFP peaks.  The LFP signal is shown together with the detected
nLFPs (circles).  Middle: nLFP-based wave-triggered average (WTA) of 
unit activity, showing that the negative peaks were associated with
an increase of neuronal firing.  Bottom: rasters of nLFP activity.
The same procedure is compared for high threshold (left panels)
and low threshold (right panels).}
\label{nLFP}
\end{figure}

We start by analyzing the power-law scaling from experimental data. To analyze
the power-law relations from LFP activity, we exploited the well-known relation
between negative LFP peaks and neuronal firing.  We \corrfirst{identified} the
negative peaks of the LFPs \corrfirst{(nLFPs), corresponding to events exceeding a
fixed threshold,} as shown in Fig.~\ref{nLFP}.  The detection was done
numerically using a fixed threshold, after digital filtering of the low-frequency
components of the signal and the detected peaks were then repositioned in the
intact original signal \corrfirst{(see Methods).}  The results of this detection for
two different thresholds are displayed in Fig.~\ref{nLFP} (top). The detected LFP
negative peaks are clearly related to neuronal firing, \corrfirst{as evidenced by the}
wave-triggered average (WTA) of the unit activity.  Indeed, the average unit
activity presented a clear increase \corrfirst{of the discharge probability} related
to the presence of negative peaks of the LFP (Fig.~\ref{nLFP}, middle).  The same
procedure was repeated for all channels, leading to rasters of nLFP activity
(Fig.~\ref{nLFP}, bottom).

\begin{figure}[ht!]
\begin{center}
 \includegraphics[width=0.8\textwidth]{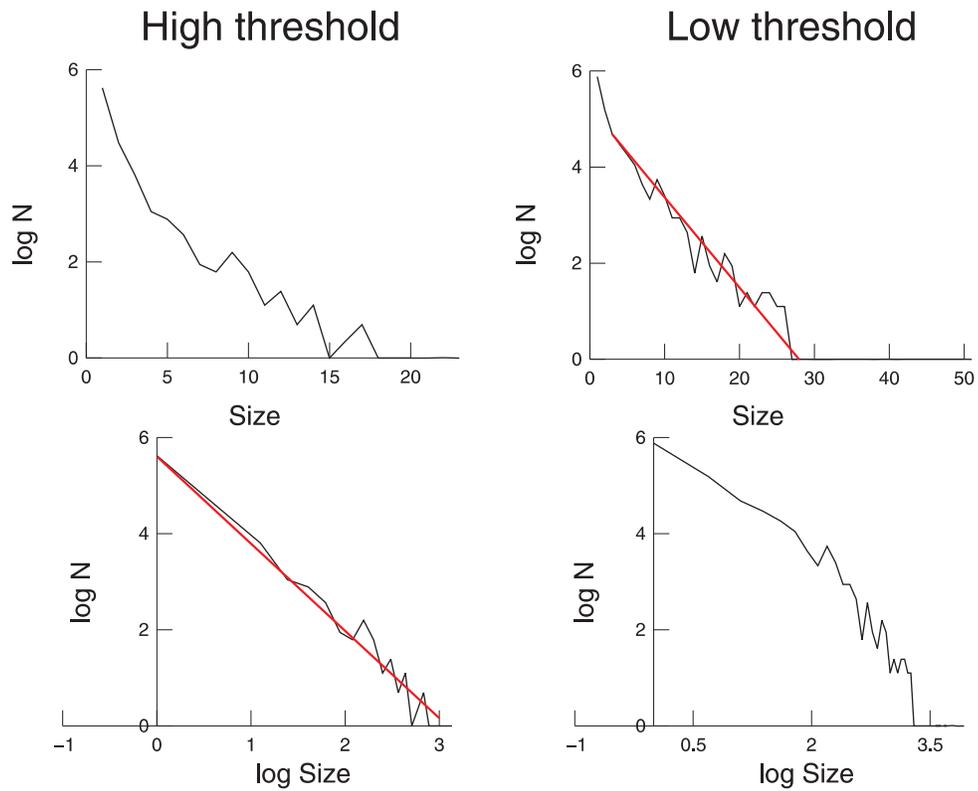}
\end{center} 
\caption{{\bf Avalanche analysis of nLFPs in the awake cat.} The nLFP
avalanche size distributions were computed according to an avalanche
analysis (see text). \corrfirst{For a high detection} threshold, the avalanche distribution is
better fit by a power-law (left panels); \corrfirst{for a low detection threshold,} 
it is better explained by an exponential distribution (right panels). }
\label{lowhigh} 
\end{figure}

We next performed an avalanche analysis based on the occurrence of nLFPs. 
\corrfirst{Similar to previous studies~\cite{beggs-plenz:04,peterman:09}, avalanches
were defined by detecting clusters of activity across all electrodes, separated
by silent periods (see Methods).  Fig.~\ref{lowhigh} shows the distribution of
avalanche size (summed amplitudes of all LFP peaks within each avalanche) in
log-linear and log-log representations and for two different detection
thresholds.}  For high threshold, the avalanche distribution was better fit by a
power-law, whereas for low threshold it \corrfirst{was better fit by an} exponential
distribution.  Similar results were obtained when the \corrfirst{avalanche} size was
defined as the total number of events (peaks) within each avalanche (not shown). 
This shows that the exact functional form of the distribution highly depends on
the peak detection threshold. Using a high detection threshold may give the
impression of a power-law relation, but lowering the threshold makes the system
tend \corrfirst{more} to an exponential distribution, consistent with the exponential
scaling of avalanches calculated from unit activity \corrfirst{in the same
experimental data~\cite{bedard-kroger-etal:06}.}

To assess the significance of this result, we performed a
Kolmogorov-\corrfirst{Smirnov} test to the same data.  The results of this test are
presented in Table~\ref{tab:gofDataAvalanches} for \corrfirst{avalanche size} defined
\corrfirst{by} the cumulated peak amplitudes.  We observe that the distribution of
\corrfirst{avalanche size} is globally well fit by an exponential distribution, which
is valid for a large proportion of the data. Indeed, an exponential fit yields
significant p-values for both low and high threshold.  Moreover, the estimated
parameters for exponential fit hardly change when the threshold is varied,
\corrfirst{again} suggesting that the observed exponential fit is meaningful. 
\corrfirst{In contrast}, the estimated power-law parameters change significantly when
changing the detection threshold, \corrfirst{and the low Kolmogorov--Smirnov distance
and high p-value obtained for low thresholds correspond to fits of only a small
percentage of the data.} Thus, although the power-law distribution seems to
provide a good fit \corrfirst{when only assessed} by a linear regression in a log-log
representation, this apparent good fit is not supported by the statistical
analysis. \corrfirst{Instead, the large negative value of the log likelihood ratio,
and the very high statistical significance of this test on these data, reveals
that the avalanche-size distribution is \corrfirst{globally} better fit by an
exponential distribution.} 

\corrfirst{The statistical avalanche analysis performed when the avalanche size was
defined as the total number of events (peaks) within each avalanche give an even
more ambiguous result. Indeed, both the exponential and the power-law
distributions provide a good fit to the data, and the log likelihood \corr{indicates that the exponential law provides a better fit} but it has a null
significance, so does not give any information on the law that best fits the data
(see Table~\ref{tab:gofEventsAvalanches})}.

\corrfirst{While these findings suggest that the the nLFP avalanches may also be
exponentially distributed, this exponential scaling may be artifactual. Although
the underlying neural activity may follow a power-law distribution, the
low-threshold condition could add spurious peaks unrelated to neuronal
activity, and that would give an exponential trend to the distribution.  This
increased ``noise'' is evident in the WTA in Fig.~\ref{nLFP}, which shows a
weaker relation to spiking activity at low threshold compared to high threshold. 
Thus, additional analyses are needed to determine which of the power-law or
exponential scaling is the more closely related to neural activity.}

\begin{figure}[ht!]
\begin{center}
 \includegraphics[width=0.8\textwidth]{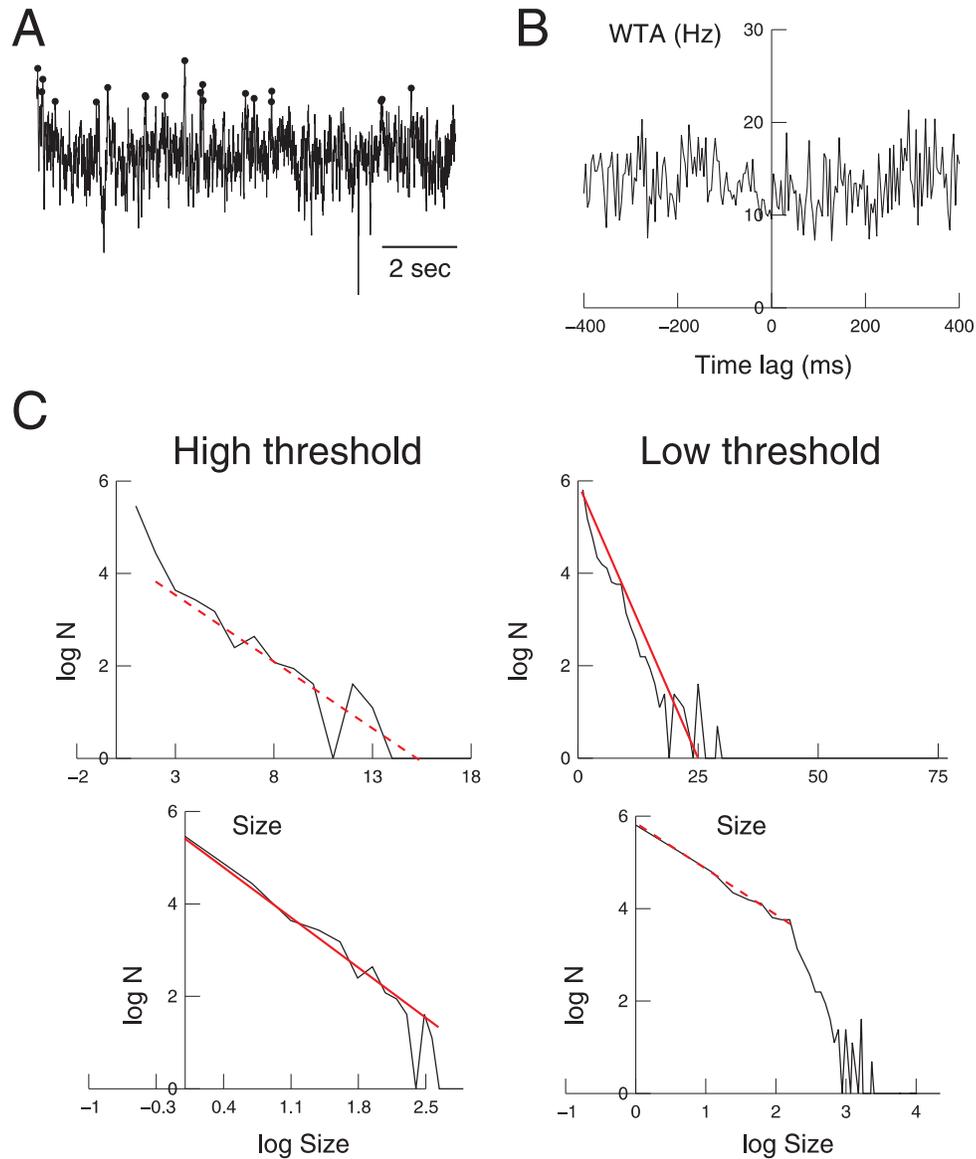}
\end{center} 
\caption{{\bf Avalanche analysis \corrfirst{of} positive LFP peaks \corrfirst{in the awake cat}.} A.  Detection
of positive LFP peaks using identical procedures as for nLFPs.  B. \corrfirst{The WTA indicates no relation between pLFPs and unit activity.}
C. Scaling of avalanche size distribution, showing similar behavior as \corrfirst{observed}
for nLFPs (compare with Fig.~\ref{lowhigh}).}
\label{positive}
\end{figure}

To further test the dependence on unit activity, we have repeated the same
avalanche analysis, but using positive peaks of the LFP \corrfirst{(pLFP;
Fig.~\ref{positive}A)}.  In this case, as expected, the peaks are not related to
unit firing (Fig.~\ref{positive}B).  Unexpectedly, however, the scaling relations
observed in graphical representations \corrfirst{for pLFPs} are similar \corrfirst{to those
observed} for nLFPs (Fig.~\ref{positive}C): \corrfirst{the low-thresholded data fits}
both \corrfirst{a} power-law and \corrfirst{an} exponential law and \corrfirst{the
high-thresholded data only fits an} exponential law.  The statistical
\corrfirst{analysis} reveals a power-law for low-threshold \corrfirst{pLFPs} and an
exponential law for high threshold pLFPs.  Interestingly, there are also some
regions where \corrfirst{both} the high \corrfirst{and low threshold pLFPs} distributions
\corrfirst{display exponential scaling} (Fig.~\ref{positive}C, dotted lines).  Here,
the Kolmogorov--\corrfirst{Smirnov} test gave results very close to the case of
negative peaks. Thus, similar to negative peaks, the apparent good fit of the
power-law distribution is not supported by the statistical analysis, \corrfirst{as
confirmed by the log likelihood ratio test.}

\begin{figure}[ht!]
\begin{center}
 \includegraphics[width=0.8\textwidth]{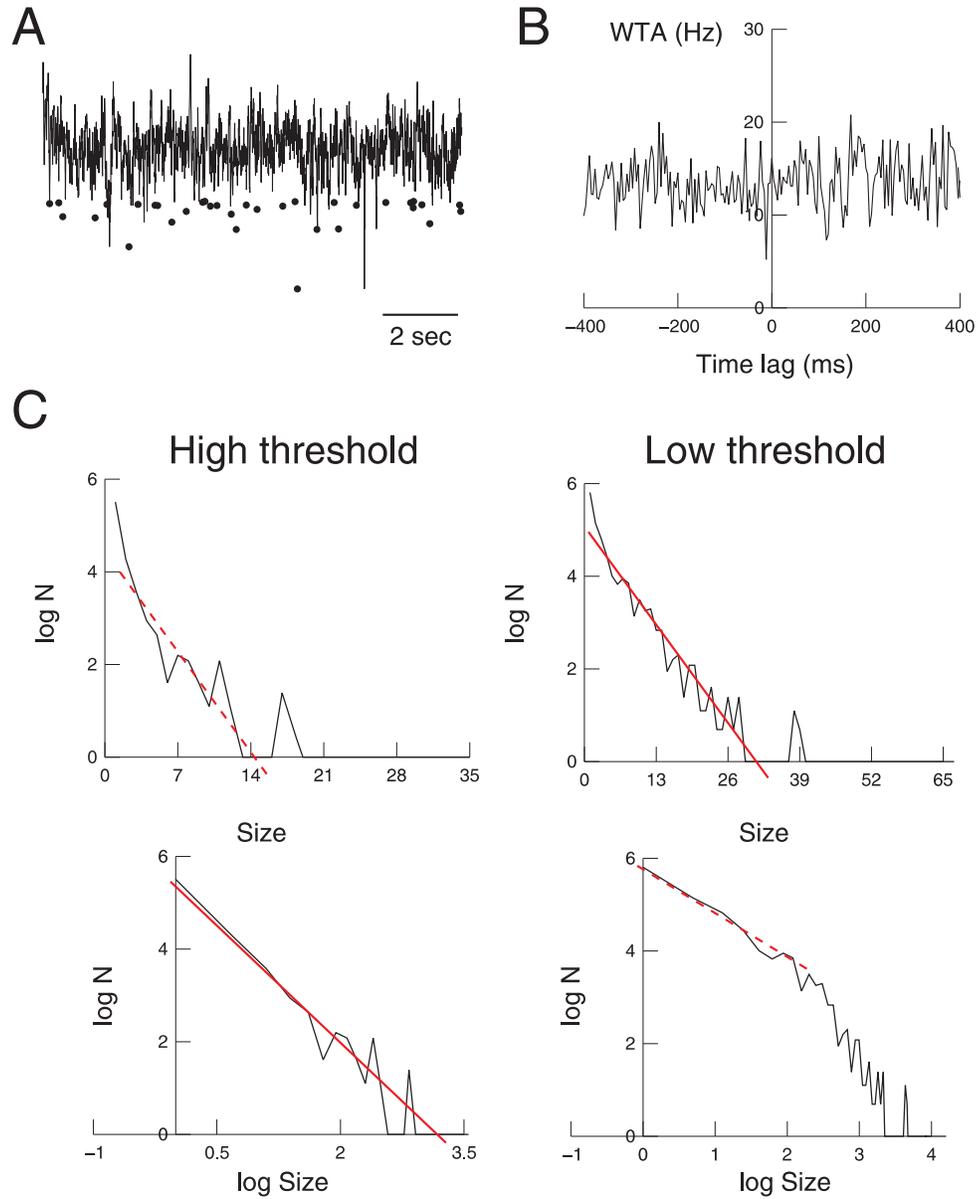}
\end{center}
\caption{{\bf Avalanche analysis \corrfirst{of} shuffled negative LFP peaks.} A.
Shuffled peaks obtained from randomizing the \corrfirst{timing} of
nLFP peaks.  B. \corrfirst{The WTA indicates that shuffling removes the relationship between nLFPs and neural activity} 
C. Scaling of avalanche peak size distribution, showing
similar behavior as for nLFPs (compare with Fig.~\ref{lowhigh}).}
\label{shuffle}
\end{figure}

Another essential test is to generate surrogate data sets.  \corrfirst{These were
produced} by taking the nLFP data sets, and randomly shuffling the occurrence
times of the different peaks, \corrfirst{while} keeping the same distribution of peak
amplitudes \corrfirst{(see Methods)}.  The avalanche analysis was then repeated
\corrfirst{using} these randomized events, and the result is shown in
Fig.~\ref{shuffle}.  \corrfirst{The shuffling ensures that} there is no correlation
between these peaks and unit activity, but interestingly, the same relations
\corrfirst{observed for the nLFPs and pLFPs} still persist.  In particular, it is
quite unexpected that \corrfirst{this} stochastic system seems to give power-law
distributed avalanche sizes.  This power-law scaling was seen mostly \corrfirst{in
the} high threshold, while the low-threshold condition behaved more
exponentially.  The opposite scaling was also seen in restricted regions
(Fig.~\ref{shuffle}C, dotted lines).  The statistical tests realized on these
surrogate data gave similar results as above (not shown).

\begin{figure}[ht!]
\begin{center}
 \includegraphics[width=0.8\textwidth]{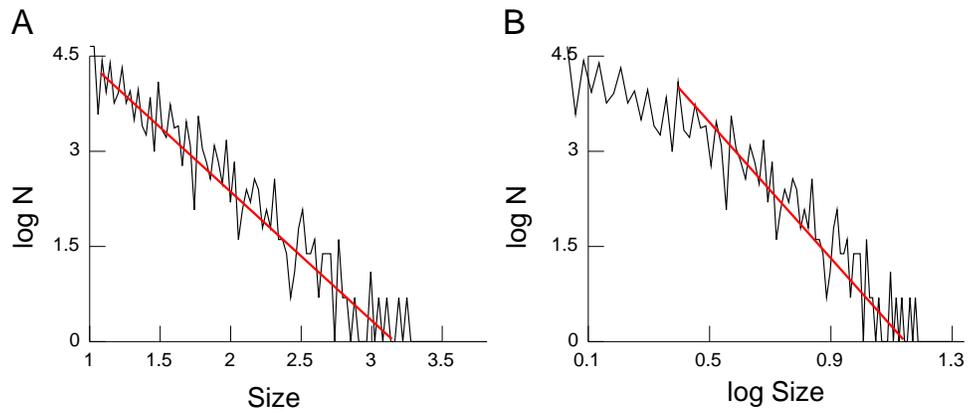}
\end{center}
\caption{{\bf \corrfirst{Avalanche-size} distributions of negative LFP peaks from 
single channels.} The peak distribution is shown \corrfirst{on} log-linear (A) and 
logarithmic scale (B).}
\label{single}
\end{figure}

The power-law scaling of nLFP size distributions was also apparent \corrfirst{{when
representing graphically the peak distributions} from single LFP channels}, as
illustrated in Fig.~\ref{single}.  To assess the significance of this result, we
performed a Kolmogorov--\corrfirst{Smirnov} test to these data (results are
\corrfirst{provided} in Table~\ref{tab:gofsingleElectrode}). \corrfirst{For most channels,
although graphically we were able to fit the data with a power-law and an
exponential distribution, the statistical tests shows that in neither case the
fit is statistically {significant}. For some channels (namely channels 1, 2 and
6), the \corrfirst{peak distribution} analysis shows, similarly to the
multi-electrodes case, that both power-law and exponential distributions provide
a good fit to the data, and the log-likelihood ratio test indicates with a high
significance level that the data are better fit by an exponential law.}

\corrfirst{These results suggest that the power-law scaling seen in log-log
representations is not necessarily} related to neuronal activity, but could
rather represent a generic property of these signals.  To test this hypothesis,
we now turn to the analysis and simulation of stochastic processes.

\subsection*{\corrfirst{Peak size distributions} from stochastic processes}

We first investigate computationally whether a power-law relation can be obtained
from the peak size distribution of a purely stochastic process.  To this end, we
generate a high-frequency shot-noise process \corrfirst{(as described in Methods)},
consisting of exponential events convolved with a Poisson process.

\begin{figure}[ht!]
	\begin{center}
 		\includegraphics[width=0.8\textwidth]{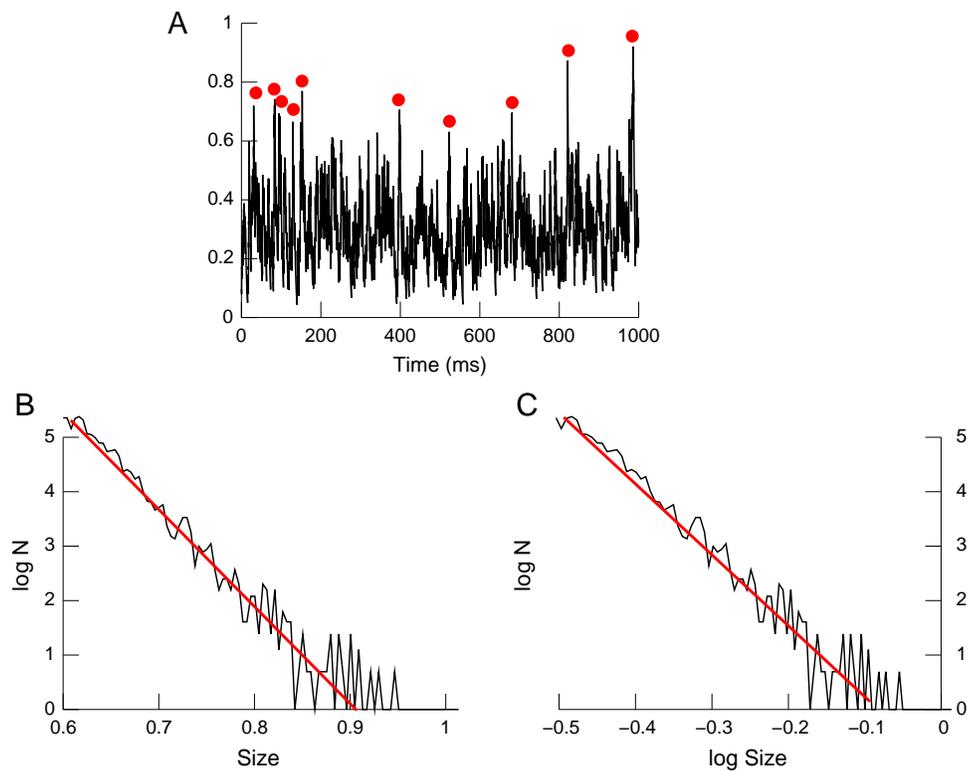}
	\end{center}
	\caption{{\bf \corrfirst{Peak-size distributions for the} thresholded Poisson shot-noise
        process.}  A. \corrfirst{Sample of the stochastic process and detected peaks.}
        B. \corrfirst{Peak size distribution on a} log-linear scale. 
        C. Same distribution \corrfirst{on a log-log} scale. Straight lines indicate
        the best fit \corrfirst{obtained} using linear regression.}
	\label{stoch} 
\end{figure}

The peaks were detected on the shot noise process $V_t$ \corrfirst{defined by
Eq.~\eqref{eq:SolShotNoise}} using a high threshold, \corrfirst{in order} to mimic the
experimental paradigm in Fig.~\ref{single}A.  As for the \corrfirst{experimental} LFP
data, this procedure yielded power-law amplitude distributions, but the same
distributions also scaled exponentially (Fig.~\ref{stoch}B-C).  

\subsubsection*{Peak distribution \corrfirst{in the} shot noise model}

We now investigate this problem analytically. \corrfirst{We treat the case where the
number of independent Poisson processes $P$ is equal or reducible to one. The
case $P>1$ can be treated in the same fashion and yields similar results. In the
case $P=1$, let us denote $t^{(i)}$ the event times of the Poisson process. T}he
integrated process \eqref{eq:SolShotNoise} simply reads:

\begin{equation}\label{eq:OnePoisson}
	V_t = V_0 e^{-t/\tau_m} + q \sum_{t^{(i)} \leq t} e^{-(t-t^{(i)})/\tau_m}
\end{equation}

We are interested in the probability that the supremum of this process reaches a
certain threshold value $\theta$ during an interval of times $[0,T]$. In order to
compute this probability, we condition on the number of jumps of the Poisson
process in this interval of time\corrfirst{, $\mathcal{N}([0,T])$}. Since the events are disconnected, we have:

\begin{align}
\nonumber \mathbbm{P} \left ( \max_{[0,T]} V_{t} \geq \theta \right) &= \sum_{N\in \mathbbm{N}} \mathbbm{P} \left ( \max_{[0,T]} V_{t} \geq \theta \cap \mathcal{N}([0,T]) = N \right)\\
\nonumber & = \sum_{N\in \mathbbm{N}} \mathbbm{P} \left ( \max_{[0,T]} V_{t} \geq \theta \big \vert \mathcal{N}([0,T]) = N \right) \mathbbm{P} \left ( \mathcal{N}([0,T]) = N \right)\\
\label{eq:ProbShotNoiseSeries} & = e^{-\lambda T}\sum_{N\in \mathbbm{N}}\frac{ ( \lambda T)^{N}}{N!} \mathbbm{P} \left ( \max_{[0,T]} V_{t} \geq \theta \big \vert \mathcal{N}([0,T]) = N \right)
\end{align}

The maxima of this process occur at the \corrfirst{event} times of the Poisson process, 
$t^{(i)}$, and have the values:
\begin{equation}\label{eq:Maximas}
\begin{cases}
t=0 & V_{0}\\
t=t^{(1)} & V_{1} := V_{0}e^{-t^{(1)}/\tau_{m}}  + q\\ 
t=t^{(2)} & V_{2} :=V_{0}e^{-t^{(2)}/\tau_{m}}  + q \left( e^{- (t^{(2)} -t^{(1)})/\tau_{m} } +1 \right) \\
\ldots\\
t=t^{(N)} & V_{N} := V_{0}e^{-t^{(N)}/\tau_{m}}  + q \Big (e^{- (t^{(N)} -t^{(1)})/\tau_{m}} + e^{- (t^{(N)} -t^{(2)})/\tau_{m}}  + \ldots \\
& \quad \qquad + e^{- (t^{(N)} -t^{(N-1)})/\tau_{m}} +1\Big)\\
& =  e^{- t^{(N)} /\tau_{m}} \left ( V_{0} + q \sum_{i=1}^{N}  e^{t^{(i)}/\tau_{m}} \right)
\end{cases}
\end{equation}

Furthermore, conditionally on $ \mathcal{N}([0,T])$ the number of jumps of the
Poisson process in the time interval $[0,T]$, the instants of these jumps are
uniformly distributed in the interval $[0,T]$. Therefore, the probability that
\corrfirst{a local maximum} is greater than the threshold $\theta$ can be
written as the following integral:
\begin{multline}
\mathbbm{P} \left ( \max_{[0,T]} V_{t} \geq \theta \big \vert \mathcal{N}([0,T]) = N \right) = \int_{t^{(1)}=0}^{T} \int_{t^{(2)}=0}^{T} \ldots \int_{t^{(N)}=0}^{T}  \\
\mathbbm{1}_{ \left\{\exists k \in \; \{1, \ldots N \} \text{ such that } \corrfirst{V_{k}} \geq \theta\right \}} \frac{dt^{(1)}\ldots dt^{(N)}}{T^{N}}
\end{multline}
where $\mathbbm{1}_{A}$ is the \corrfirst{indicator} function of the set $A$. 
Therefore, the peak distribution we are searching for has the expression:
\begin{multline}\label{eq:DistribShotNoiseSerie}
\mathbbm{P} \left ( \max_{[0,T]} V_{t} \geq \theta \right) = e^{-\lambda T}\sum_{N\in \mathbbm{N}}\frac{ ( \lambda T)^{N}}{N!}
\int_{t^{(1)}=0}^{T} \int_{t^{(2)}=0}^{T} \ldots \int_{t^{(N)}=0}^{T} \\
\mathbbm{1}_{ \left \{ \exists k \in \; \{1, \ldots N \} \text{ such that } \corrfirst{V_{k}} \geq \theta \right \}} \frac{dt^{(1)}\ldots dt^{(N)}}{T^{N}}
\end{multline}
\corrfirst{This integral cannot be simplified further, but can be accurately
approximated using a numerical integration method and truncating the series. }
The approximation error is proportional to the rest of the exponential series
$R(N) = \sum_{k=N+1}^{\infty} (\lambda T)^{k}/k!$.

Let us now consider the distribution of the maxima of the process
\eqref{eq:OnePoisson} \corrfirst{given that} the process does an excursion above a
certain threshold. This case can be treated in a similar fashion, but considering
the distribution of local minima also. These local minima are reached at times
$t^{(k) -}$ just before the jumps of the Poisson process, and their value are
$V_{t^{(k)}}-q$.  The probability of an excursion above $\theta$ and exceeding
$\mu$ (event denoted $A_{\theta}^{\mu}$) can therefore be written as:

\begin{align*}
\mathbbm{P} \left ( A_{\theta}^{\mu} \right) &= \sum_{N\in \mathbbm{N}} \mathbbm{P} \left ( A^{\mu}_{\theta} \cap \mathcal{N}([0,T]) = N \right)\\
& = e^{-\lambda T}\sum_{N\in \mathbbm{N}}\frac{ ( \lambda T)^{N}}{N!} \mathbbm{P} \left ( A^{\mu}_{\theta}  \big \vert \mathcal{N}([0,T]) = N \right)
\end{align*}
\noindent and the probability $\mathbbm{P} \left ( A^{\mu}_{\theta}  \big \vert \mathcal{N}([0,T]) = N \right)$ can be easily evaluated numerically using the following representation:
\begin{multline}
\mathbbm{P} \left ( A_{\theta}^{\mu} \big \vert \mathcal{N}([0,T]) = N \right) = \int_{t^{(1)}=0}^{T} \int_{t^{(2)}=0}^{T} \ldots \int_{t^{(N)}=0}^{T}  \\
\mathbbm{1}_{ \corrfirst{\{} \exists k \in \{1,\ldots N\} \text{ and } l \in  \{k+1, \ldots N\} \text{ such that } V_{k} \geq \mu, \; V_{l} \leq \theta \corrfirst{\}} } \frac{dt^{(1)}\ldots dt^{(N)}}{T^{N}}
\end{multline}

\begin{figure}[ht!]
\begin{center}
\includegraphics[width=.9\textwidth]{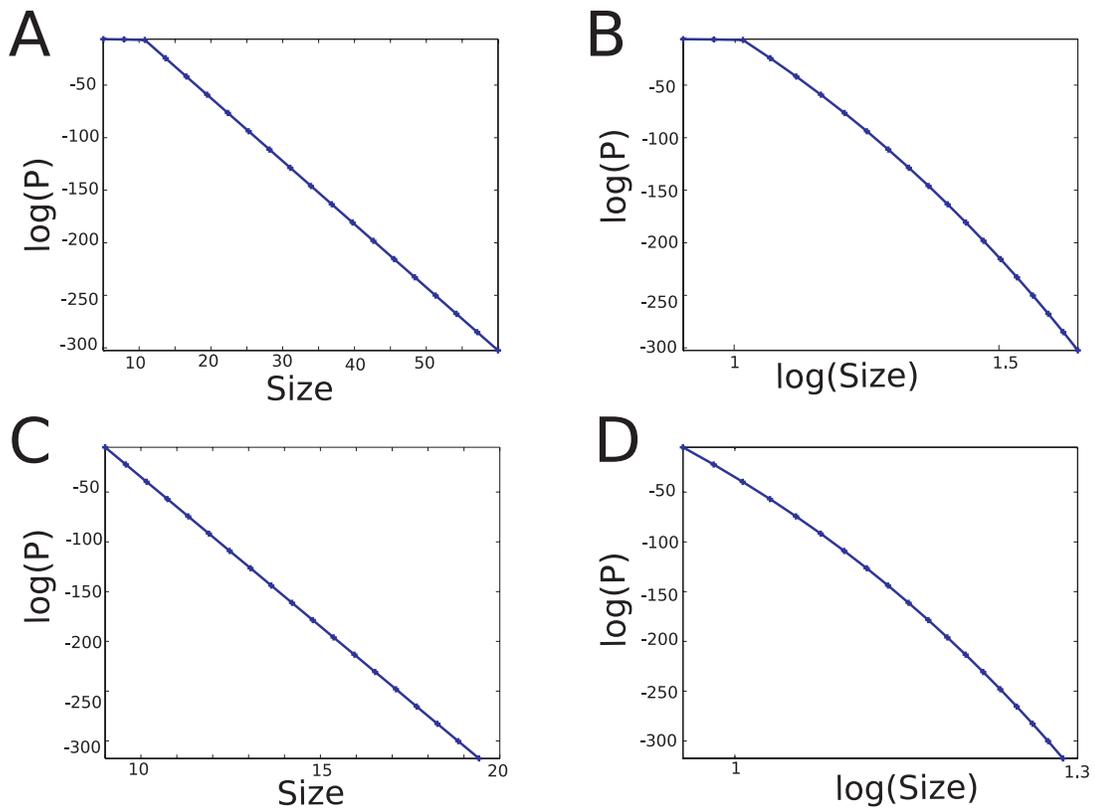}
\end{center}
\caption{{\bf Peak amplitude distribution for the Shot-Noise model.} 
\corrfirst{Single-barrier case (A,B) on a log-linear scale (A) and on a log-log scale 
(B) show a globally linear trend. Excursions (C,D) show exactly the same 
profile.} Simulation \corrfirst{parameters}: intensity 
of the process $\lambda=4$, $\tau_{m}=2$, $V_{0}=0$, $T=10$, 
$\theta=10$, maximal value of peaks considered $25$ (see text)}
\label{fig:resultsTheoSN}
\end{figure}

\corrfirst{Simulation results of these distributions are presented in
Fig.~\ref{fig:resultsTheoSN} and predict the results obtained by numerical
simulations in Fig.~\ref{stoch}: both exponential and power-law distributions
give a good model for the peak amplitude distribution. The results of the
statistical analysis are in accordance with this observation, and are provided in
Table~\ref{tab:TheoreticalLaws}. Indeed, we observe that the exponential 
\corrfirst{distribution gives a good model in both the single barrier and the excursion case, and} the power-law distributions provide a good agreement with the computed
theoretical distributions \corrfirst{only in the excursion case}. Note that we did not use the log-likelihood ratio
because this statistical test is defined through the computation of the
likelihood of a given probabilistic model on a data set, and here we do not have
data sets but we directly compute the probability distributions.}

\subsubsection*{\corrfirst{Peak distribution in the Ornstein-Uhlenbeck model}}\label{sec:DifApp}

\corrfirst{In the case of the Ornstein-Uhlenbeck model, the stochastic process
modelling the LFP} has the same regularity as the Brownian motion, \corrfirst{and
therefore is} is nowhere differentiable, and has a dense countable set of local
maxima. \corrfirst{In that case, peaks are no more defined as local maxima of the
process, and} the problem is reduced to \corrfirst{determining} the probability that the
process exceeds a certain value. This probability can be deduced from the law of
the first hitting time of the Ornstein--Uhlenbeck process. Indeed, let us denote
by $\tau_a$ the first hitting time of the threshold $a$ for the
Ornstein-Uhlenbeck process given by equation \eqref{eq:SolOU}. The probability
\corrfirst{that the process exceeds} a certain level $a$, \corrfirst{given that it
exceeds the threshold $\theta$, is given by}:
\begin{align}
\nonumber \mathbb{P} \left (\sup\limits_{s\in [0,t]} V_s \geq a \Big \vert \sup\limits_{s\in [0,t]} V_s \geq \theta , V_{0}\right ) & = \mathbb{P} \left (\sup\limits_{s\in [0,t]} V_s \geq a \Big \vert V_{0}\right) \\
\label{eq:ThreshHT} & = \mathbb{P} \left (\tau_a \leq t  \Big \vert V_{0} \right) 
\end{align}

The excursion case continuous equivalent consists in considering the
probability of exceeding a certain quantity $a$ before going back
under the excursion threshold $\theta$. This probability can be
written as:
\begin{align}
\nonumber \mathbb{P} \left (\sup\limits_{s\in [0,t]} V_s \geq a, \inf_{t\in [\tau_{a}, t]} V_{s} \leq \theta \Big \vert \sup\limits_{s\in [0,t]} V_s \geq \theta , V_{0}\right ) &= \mathbb{P} \left (\sup\limits_{s\in [0,t]} V_s  \geq a , \inf_{t\in [\tau_{a}, t]} V_{s} \leq \theta \Big \vert V_{0}\right) \\
\label{eq:ExcurHT} & = \int_{s=0}^{t} \mathbb{P} \left (\tau_{\theta} \leq t  \Big \vert V_{s}=a\right) \mathbb{P} \left (\tau_a \in ds  \Big \vert V_{0} \right)
\end{align}

Therefore, the repartition function of the maxima, and \corrfirst{that} of \corrfirst{the}
maxima above a certain threshold, can be deduced from the repartition function of
the first hitting time of the process $V$. As reviewed in
\cite{alili-patie-etal:05,touboul-faugeras:07b}, there is no closed form solution
for the probability distribution of these hitting times, but they can be
efficiently numerically \corrfirst{computed}. The most convenient solution
\corrfirst{involves} solving a Volterra integral equation \corrfirst{to obtain} \corrfirst{the
law of the first hitting time variable} (see e.g.
\cite{schrodinger:15,plesser:99, touboul-faugeras:07b}).

\begin{figure}[ht!]
\begin{center}
\includegraphics[width=.9\textwidth]{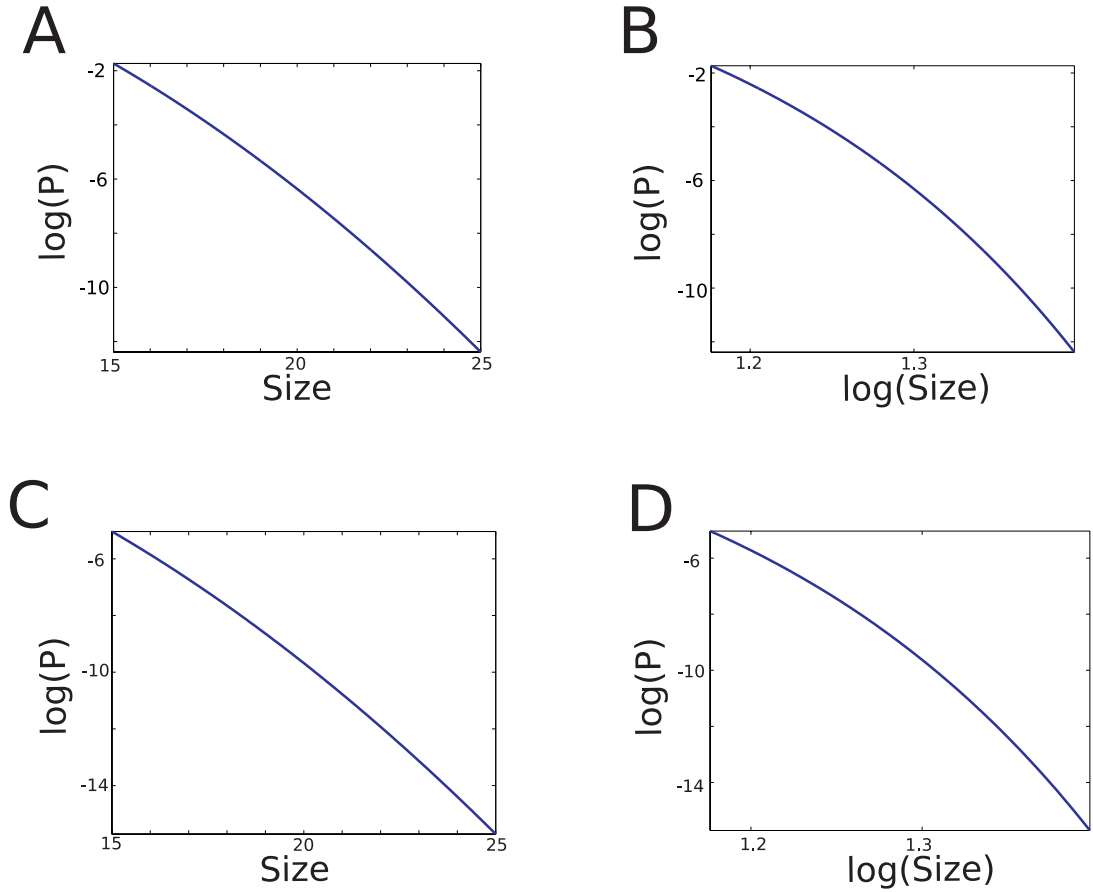}
\end{center}
\caption{{\bf Peak amplitude distribution \corrfirst{for the Ornstein-Uhlenbeck process}}. \corrfirst{(A,B) : single-barrier peaks, on a log-linear scale (A) and on a log-log 
scale (B), and excursions (C,D), on a log-linear scale (C) and on a log-log scale (D). Both case present the same profile and a globally linear trend for both axis.} Simulation \corrfirst{parameters}: intensity of the process $\lambda=4$, 
$\tau_{m}=2$, $V_{0}=0$, $T=10$, $\theta=10$, maximal value of peaks 
considered: $25$ \corrfirst{(see text)}.}
\label{fig:resultstheoOU}
\end{figure}

\corrfirst{In this case again, the same remarks apply: we observe (see
Fig.~\ref{fig:resultstheoOU}) for \corrfirst{both} the single-barrier \corrfirst{and the
excursion problems} that the peak-amplitude distribution is fit equally well by
either a power-law or exponential distribution. This is supported by the more
rigorous statistical analysis (see Table~\ref{tab:TheoreticalLaws}): both the
exponential and the power-law distributions provide a good agreement with the
distributions computed numerically form the formulas derived. }

\corrfirst{\subsection*{Avalanche size distribution in a neural network at
criticality}}

\begin{figure}[ht!]
\begin{center}
\includegraphics[width=.9\textwidth]{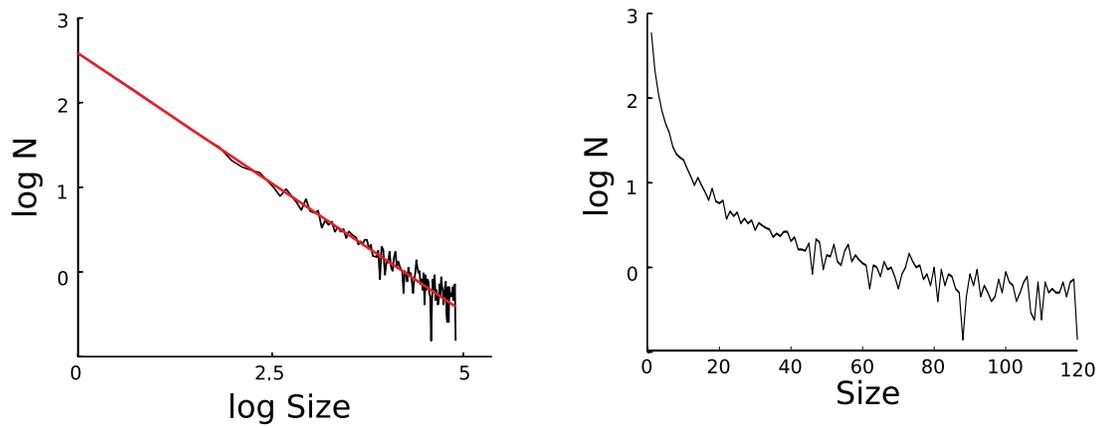}
\end{center}
\caption{ \corrfirst{{\bf Avalanche analysis of a simulated neural network 
displaying self-organized criticality}. The power-law distribution provides a 
very good graphical fit \corrfirst{(A)}, whereas
the exponential distribution provides a poor fit \corrfirst{(B)}.  Data from
ref.~\cite{levina-etal:07}} }

\label{fig:AnnaLevina}
\end{figure}

\corrfirst{We finally performed the above statistical analysis on the avalanche data
generated by the artificial network in the critical state of Levina and
colleagues \cite{levina-etal:09,levina-etal:07} (data kindly provided by Anna
Levina).  The avalanche size distributions obtained are plotted in
Fig.~\ref{fig:AnnaLevina}, and the results of the statistical analysis show very
clearly that the data are very well fitted by a power-law in this case (see
Table~\ref{tab:Datanna}). We conclude that in the case of a neural network at
criticality, the ambiguity observed in the experimental data is not present, even
when using the same number of avalanches as in our data.  Thus, this analysis
brings another argument to support the absence of robust power-law scaling in the
experimental data.}

\

\section*{Discussion}

In this paper, we have provided an analysis of multisite LFP recordings in awake
cats, \corrfirst{using} the detection of negative LFP peaks (nLFPs), as done in a
previous study~\cite{peterman:09}.  The analysis shows that the
\corrfirst{occurrence time} and amplitudes of nLFPs can show power-law distributions,
\corrfirst{but in a manner that depends on the detection threshold.  High thresholds,
which select events of exceptionally large amplitude, tend to give power-law
relations.  In contrast, low thresholds, which select many events, give rise to
exponential distributions, similar to stochastic processes. The application of
more \corrfirst{rigorous} statistical tests, such as the Kolmogorov--\corrfirst{Smirnov}
test, shows that the power-law relations are not supported \corrfirst{by} solid
statistical grounds.  The dependence on the threshold is much weaker in the
statistical data analysis, as we can clearly see in
Tables~\ref{tab:gofDataAvalanches} and \ref{tab:gofEventsAvalanches}.}

\corrfirst{Because the exponential scaling could be interpreted as a spurious result
due to the addition of a large number of peaks unrelated to neuronal activity, we
considered two controls: first, positive LFP peaks, which are not related to
neuronal activity, and randomly shuffled peak times, which makes the system
equivalent to a stochastic process \corrfirst{with the same} peak amplitude
distribution \corrfirst{as the data}.  The two cases show similar apparent power-law
scaling and dependency to threshold as for nLFPs.}

\corrfirst{These results suggest that the spurious power-law scaling could be a
generic property of thresholded stochastic processes.  To investigate this point
in more depth, we studied a similar peak detection paradigm applied to two simple
stochastic models,} one corresponding to LFPs arising from a linear summation of
spikes arriving at the times of a Poisson process (a shot-noise process) and the
diffusion limit of this phenomenon \corrfirst{(an Ornstein--Uhlenbeck process)}.  The
former case can be solved in a closed \corrfirst{integral} form while the latter case
is solved using the laws of the first hitting times of the Ornstein-Uhlenbeck
process. Both \corrfirst{models demonstrate the same ambiguity:} when only looking at
the log-linear and log-log plots, and both power-laws and exponential laws can be
fitted.  However, the application of \corrfirst{the} more \corrfirst{rigorous}
Kolmogorov--\corrfirst{Smirnov} test demonstrated that some apparent power-law scaling
(as seen from log-log representations) is not \corrfirst{based on} solid statistical
grounds, in real data as well as in the theoretical laws computed, in agreement
with previous studies (see e.g.~\cite{clauset-etal:09}).

\corrfirst{This analysis therefore confirms that thresholded stochastic processes can
display power-law scaling, but only when performing simple line fitting in
log-log representations. Indeed, we observe that it is always possible to fit a
power-law distribution to the tail of the distribution with a quite good
agreement, but these fits do not hold for large \corrfirst{threshold values} (see
Table~\ref{tab:TheoreticalLaws}).  The estimated laws \corrfirst{yielded} high values
of the exponent which is not very realistic in general. This is consistent with
the findings reported above for LFPs: the power-law scaling of LFP peaks displays
very similar properties to that of stochastic processes, which supports the idea
that \corrfirst{experimentally observed} power-law scaling is not \corrfirst{necessarily}
related to neuronal activity, but may be explained by a generic property of
thresholded stochastic processes.}

\corrfirst{The same analysis applied to a network presenting self-organized
criticality confirms with no ambiguity that the distribution of avalanche size
presents a clear power-law distribution, whereas in cortical LFPs the
power-law scaling in log-log representations was not supported by statistical
analyses. We conclude that power-law scaling, particularly when deduced from
log-log representations, does not constitute a proof of self-organized
criticality, but should be complemented by more sophisticated statistical
analyses.}

Thus, contrary to a previous study in monkey~\cite{peterman:09}, where the same
controls were not done, our analysis suggests that, in awake cats, the
power-law scaling is not related to neuronal activity \corrfirst{but is rather
an artefact of the thresholding procedure.  In agreement with this,} a
previous analysis~\cite{bedard-kroger-etal:06} failed to see evidence for
power-law distributions and avalanche dynamics from spiking activity in the same
data set, which rather scaled exponentially.  However, there is still the
possibility that \corrfirst{these differences arise from other factors such as the
different species, brain areas, or cortical layers used in these experiments.}. 
Further studies should address these points.

\section*{Acknowledgments}

Research supported by the NSF (grant DMS0817131), CNRS, ANR and the European
Community (FACETS project).  \corrfirst{We thank Anna Levina and colleagues for kindly
providing us avalanche data from their neural network model published in
\cite{levina-etal:07}, and John Agapiou for comments on the manuscript.}

\clearpage

\section*{Tables}

\begin{table}[!ht]
	\caption{\bf \corrfirst{Results of avalanche-analysis (summed LFP amplitudes)}}
	\begin{center}
	\begin{tabular}{|c||c|c|c|c||c|c|c|c||c|c|c|}
		\hline
		Data Type  & \multicolumn{4}{|c||}{Exponential fit} & \multicolumn{4}{|c||}{Power-Law fit} & \multicolumn{3}{|c|}{\corrfirst{Log Likelihood}}\\
	 	and threshold & \multicolumn{4}{|c||}{}& \multicolumn{4}{|c||}{} & \multicolumn{3}{|c|}{\corrfirst{ratio}}\\
		\hline
	  & $\lambda$ & \corrfirst{KS} & p-val & \% & $\alpha$ & \corrfirst{KS} & p-val & \% & \corr{LLR} & \corr{p-val} & \corr{Result}\\
		\hline
		Neg. Low & .18 & 0.028 & 0.07 & 38 & 5.32 & 0.050 & 0.94 & 4 & \corr{-1211} & \corr{0.0} &\corrfirst{Exp}\\
		\hline
		Neg. High & 0.13 & 0.042 & 0.07 & 20 & 2.01 & 0.077 & 0 & 88 & \corr{-133} & \corr{0.0} & \corrfirst{Exp}\\
		\hline
		Pos. Low & 0.079 & 0.052 & 0 & 2.7 & 2.97 & 0.041 & 0.70 & 9 & \corr{-1275} & \corr{0.0} & \corrfirst{Exp}\\
		\hline
		Pos High & 0.18 & 0.033 & 0.27 & 31 & 1.93 & 0.091 & 0 & 93 & \corr{-351} & \corr{0.0} & \corrfirst{Exp}\\
		\hline
	\end{tabular}
	\end{center}
  \begin{flushleft}
        \corrfirst{Results of avalanche analysis performed on data obtained from } 
positive (Pos) and negative (Neg) LFP peaks detected with \corrfirst{either} a Low or High
threshold. In this analysis, the avalanche size was the \corrfirst{summed} amplitude
of all LFP peaks within the avalanche. The $\lambda$ is the estimated exponential
parameter, the \corrfirst{KS value corresponds to the Kolmogorov--Smirnov distance}.  The smaller the \corrfirst{KS}, the better
the fit.  The closer to 1 the p-value, the better the fit.  The \% represents the
percentage of data explained by the best fit.  We observe that the distributions
of these data \corrfirst{are} better represented by exponential fits, and the positive peaks
with a low threshold are not well modeled by \corrfirst{either an exponential law or a
power law}. \corrfirst{The log-likelihood ratio test always conclude that a better 
fit is provided by the exponential law. This test is performed \corrfirst{over a common range by fitting the data using the same values of $x_{\min}$ and $x_{\max}$ for both laws}.
\corr{The \emph{LLR} value is the value of the log-likelihood ratio. It is negative (positive) if the better fit is the exponential (power-law) distribution. The \emph{p-val} value is the significance log-likelihood ratio (see text). The closer to 0, the more significant the test. \emph{Result} corresponds to the conclusion of the log likelihood  ratio test:} \emph{Exp} indicates that the log likelihood ratio concludes that the exponential fit is better.}
\end{flushleft}
	\label{tab:gofDataAvalanches}
\end{table}

\begin{table}[!ht]
	\caption{\bf \corrfirst{Results of the avalanche size analysis (number of LFP peaks).}}
	\begin{center}
	\begin{tabular}{|c||c|c|c|c||c|c|c|c||c|c|c|}
		\hline
		Data Type  & \multicolumn{4}{|c||}{Exponential fit} & \multicolumn{4}{|c||}{Power-Law fit} & \multicolumn{3}{|c|}{\corrfirst{Log-Likelihood}}\\
	 	and threshold & \multicolumn{4}{|c||}{}& \multicolumn{4}{|c||}{} & \multicolumn{3}{|c|}{\corrfirst{ratio}}\\
		\hline
	  & $\lambda$ & \corrfirst{KS}  & p-val & \% & $\alpha$ & \corrfirst{KS} & p-val  & \% & \corr{LLR} & \corr{p-val} & \corr{Result}\\
		\hline
			Neg. Low  & \corrfirst{0.19} & \corrfirst{0.023} & \corrfirst{0.64} & \corrfirst{54} &\corrfirst{1.26} & \corrfirst{0.020} & \corrfirst{0.83} & \corrfirst{18} & \corr{-77} & \corr{1.0} & \corrfirst{ $\star$ (Exp)  }\\
			\hline 
			Neg. High & \corrfirst{0.27} & \corrfirst{0.045} & \corrfirst{0.27} & \corrfirst{29} & \corrfirst{1.74} & \corrfirst{0.009} & \corrfirst{0.97} & \corrfirst{100} & \corr{-61} & \corr{1.0} & \corrfirst{ $\star$ (Exp) }\\
			\hline 
			Pos Low & \corrfirst{0.23} & \corrfirst{0.030} & \corrfirst{0.19} & \corrfirst{70} & \corrfirst{1.20} & \corrfirst{0.021} & \corrfirst{0.60} & \corrfirst{54} & \corr{-232} & \corr{1.0} &\corrfirst{ $\star$ (Exp)}\\ 
			\hline 
			Pos. High & \corrfirst{0.36} & \corrfirst{0.067} & \corrfirst{0.14} & \corrfirst{50} & \corrfirst{1.54} & \corrfirst{0.012} & \corrfirst{0.91} & \corrfirst{100} & \corr{-110} & \corr{1.0} & \corrfirst{ $\star$ (Exp)}\\
			\hline 
	\end{tabular}
        \end{center}

	\begin{flushleft}
		{\corrfirst{Results of avalanche analysis for avalanche size defined as the number of LFP} peaks within the avalanche, for both positive and
	negative events. \corrfirst{Table headers are the same as in }
	Table~\ref{tab:gofDataAvalanches}. \corrfirst{ The $\star$ indicates that 
	the fit is not statistically significant.}}
	\end{flushleft}
	\label{tab:gofEventsAvalanches}
\end{table}

\begin{table}[!ht]
	\caption{\bf \corrfirst{Results of avalanche-analysis} for single-electrode LFP
peaks}
	\begin{center}
	\begin{tabular}{|c||c|c|c|c||c|c|c|c||c|c|c|}
		\hline
		Data Type  & \multicolumn{4}{|c||}{Exponential fit} & \multicolumn{4}{|c||}{Power-Law fit} & \multicolumn{3}{|c|}{\corrfirst{Log-Likelihood}}\\
	 	and threshold & \multicolumn{4}{|c||}{}& \multicolumn{4}{|c||}{} & \multicolumn{3}{|c|}{\corrfirst{ratio}}\\
		\hline
		& $\lambda$ & \corrfirst{KS}  & p-val & \% & $\alpha$ & \corrfirst{KS} & p-val & \% & \corr{LLR} & \corr{p-val} & \corr{Result}\\
	  \hline
		\hline
		Neg. Low & 2.39 & 0.029 & 0.055 & 39 & 6.17 & 0.056 & 0.00 & 33 & \corr{-47} & \corr{0.0} & \corrfirst{Exp}\\
		\hline
		Neg. High & 2.82 & 0.030 & 0.68 & 80 & 9.05 & 0.048 & 0.53 & 34 & \corr{-4.4} & \corr{0.04}  & \corrfirst{Exp}\\ 
		\hline
		Pos. Low & 2.07 & 0.022 & 0.25 & 98 & 6.15 & 0.041 & 0.06 & 26 & \corr{-37.7} & \corr{0.0}  & \corrfirst{Exp}\\
		\hline
		Pos. High & 2.21 & 0.038 & 0.29 & 56 & 6.85 & 0.044 & 0.10 & 100 & \corr{-1.29} & \corr{0.66}  & \corrfirst{$\star$ (Exp)}\\
		\hline 
	\end{tabular}
        \end{center}
	\begin{flushleft}
		\corrfirst{Results of avalanche-analysis for avalanches defined from single-electrode LFP peaks, positive and negative, with low and high threshold. Table headers are the same as in}	Table~\ref{tab:gofDataAvalanches}.\end{flushleft}
	\label{tab:gofsingleElectrode}
\end{table}

\begin{table}[!ht]
	\caption{\bf \corrfirst{Results of avalanche-analysis} for thresholded stochastic processes}
	\begin{center}
	\begin{tabular}	{|c||c|c|c||c|c|c||}
			\hline
			Data type  & \multicolumn{3}{|c|}{Exponential fit} & \multicolumn{3}{|c|}{Power-Law fit}\\
			\hline
		  & $\lambda$ & \corrfirst{KS} & p-val & $\alpha$ & \corrfirst{KS} & p-val \\
			\hline
			Shot-Noise & 0.70 & 0.103 & 0.12 & 10.08 & 0.185 & 0.00\\
			single-barrier & & & & & & \\
			\hline
			Shot-Noise &  0.72 & 0.014 & 1.00 & 15.00 & 0.094 & 0.28 \\
			excursion & & & & & & \\
			\hline
			Ornstein--Uhlenbeck & 2.40 & 0.042 & 0.97 & 44 & 0.077 & 0.62  \\
			single-barrier & & & & & & \\
			\hline
			Ornstein--Uhlenbeck &  2.42 & 0.0051 & 1.00 & 48.00 & 0.012 & 0.92 \\
			excursion & & & & & & \\
			\hline
	\end{tabular}
	\end{center}
	\begin{flushleft}
                {\corrfirst{Results of avalanche-analysis for avalanche-sizes analytically determined for four stochastic processes. Table headers are the same as in Table~\ref{tab:gofDataAvalanches}.} 
                \corrfirst{The estimated power-law is large} because we considered the tail of the distribution, and since the data present an exponential trend, the estimated power-law exponent becomes larger when thresholds are high. Even if the p-value is 
		high, the fit is not realistic and the does not hold for larger intervals. \corrfirst{We do not use the log-likelihood 
		ratio since it is defined for samples and does not really make sense for distributions.}}
	\end{flushleft}
	\label{tab:TheoreticalLaws}
\end{table}

\begin{table}[!ht]
	\caption{\bf \corrfirst{ \corrfirst{Results of avalanche-analysis}  for the artificial network model \cite{levina-etal:07} at criticality.}}
	\begin{center}
	\begin{tabular}	{|c||c|c|c||c|c|c||c|c|c|}
			\hline
			\corrfirst{
			Data type}  & \multicolumn{3}{|c|}{\corrfirst{Exponential fit}} & \multicolumn{3}{|c|}{\corrfirst{Power-Law fit}} & \multicolumn{3}{|c|}{\corrfirst{Log-Likelihood ratio}}\\
			\hline
		  & \corrfirst{$\lambda$} & \corrfirst{KS} & \corrfirst{p-val} & \corrfirst{$\alpha$} & \corrfirst{KS} & \corrfirst{p-val} & \corr{LLR} & \corr{p-val} & \corr{Result}\\
			\hline
			\corrfirst{Full data set} & \corrfirst{0.10} & \corrfirst{0.2820} & \corrfirst{0.00} & \corrfirst{1.44} & \corrfirst{0.0027} & \corrfirst{0.85} & \corr{1645} & \corr{0.0} & \corrfirst{PL} \\
			\hline
			\corrfirst{$2,000$ avalanches} &  \corrfirst{0.10} & \corrfirst{0.2806} & \corrfirst{0.00} & \corrfirst{1.42} & \corrfirst{0.0061} & \corrfirst{0.80} & \corr{2483} & \corr{0.0} & \corrfirst{PL} \\
			\hline
	\end{tabular}
	\end{center}
	\begin{flushleft}
                {\corrfirst{Results of avalanche-analysis for avalanche-sizes determined using 
                a sequence of $20,000$ avalanches produced by the artificial neural network model, 
                and a smaller set of $2,000$ avalanches corresponding to the typical number of avalanches we have in our experimental simulations. The power-law model provides a very good fit, with high p-value, 
                whereas an exponential law is not a good statistical model of the data in either case.}}
	\end{flushleft}
	\label{tab:Datanna}
\end{table}

\clearpage

\section*{Figures}

\begin{thebibliography}{10}
\providecommand{\url}[1]{\texttt{#1}}
\providecommand{\urlprefix}{URL }
\expandafter\ifx\csname urlstyle\endcsname\relax
  \providecommand{\doi}[1]{doi:\discretionary{}{}{}#1}\else
  \providecommand{\doi}{doi:\discretionary{}{}{}\begingroup
  \urlstyle{rm}\Url}\fi
\providecommand{\bibAnnoteFile}[1]{%
  \IfFileExists{#1}{\begin{quotation}\noindent\textsc{Key:} #1\\
  \textsc{Annotation:}\ \input{#1}\end{quotation}}{}}
\providecommand{\bibAnnote}[2]{%
  \begin{quotation}\noindent\textsc{Key:} #1\\
  \textsc{Annotation:}\ #2\end{quotation}}
\providecommand{\eprint}[2][]{\url{#2}}

\bibitem{bak:96}
Bak P (1996) How nature works: the science of self-organized criticality.
\newblock Copernicus New York.
\input{bak:96}

\bibitem{jensen:98}
Jensen H (1998) Self-organized criticality: emergent complex behavior in
  physical and biological systems.
\newblock Cambridge Univ Pr.
\input{jensen:98}

\bibitem{beggs-plenz:04}
Beggs JM, Plenz D (2004) Neuronal avalanches are diverse and precise activity
  patterns that are stable for many hours in cortical slice cultures.
\newblock J Neurosci 24: 5216--5229.
\input{beggs-plenz:04}

\bibitem{hennig:09}
Hennig M, Adams C, Willshaw D, Sernagor E (2009) Early-stage waves in the
  retinal network emerge close to a critical state transition between local and
  global functional connectivity.
\newblock Journal of Neuroscience 29.
\input{hennig:09}

\bibitem{steriade:01}
Steriade M (2001) Impact of network activities on neuronal properties in
  corticothalamic systems.
\newblock Journal of Neurophysiology 86.
\input{steriade:01}

\bibitem{bedard-kroger-etal:06}
Bedard C, Kroger H, Destexhe A (2006) Model of low-pass filtering of local
  field potentials in brain tissue.
\newblock Phys Rev E Stat Nonlin Soft Matter Phys 73: 051911.
\input{bedard-kroger-etal:06}

\bibitem{peterman:09}
Petermann T, Thiagarajan T, Lebedev M, Nicolelis M, Chialvo D, et~al. (2009)
  Spontaneous cortical activity in awake monkeys composed of neuronal
  avalanches.
\newblock Proceedings of the National Academy of Sciences 106.
\input{peterman:09}

\bibitem{levina-etal:07}
Levina A, Herrmann JM, Geisel T (2007) Dynamical synapses causing
  self-organized criticality in neural networks.
\newblock Nature Physics 3: 857-860.
\input{levina-etal:07}

\bibitem{levina-etal:09}
Levina A, Herrmann JM, Geisel T (2009) Phase transitions towards criticality in
  a neural system with adaptive interactions.
\newblock Phys Rev Lett 102.
\input{levina-etal:09}

\bibitem{destexhe-contreras-etal:99}
Destexhe A, Contreras D, Steriade M (1999) Spatiotemporal analysis of local
  field potentials and unit discharges in cat cerebral cortex during natural
  wake and sleep states.
\newblock J Neurosci 19: 4595--4608.
\input{destexhe-contreras-etal:99}

\bibitem{billingsley:99}
Billingsley P (1999) Convergence of Probability Measures.
\newblock Wiley series in probability and statistics.
\input{billingsley:99}

\bibitem{touboul-faugeras:07b}
Touboul J, Faugeras O (2007) The spikes trains probability distributions: a
  stochastic calculus approach.
\newblock Journal of Physiology, Paris 101/1-3: 78--98.
\bibAnnote{touboul-faugeras:07b}{ftp://ftp-sop.inria.fr/odyssee/Publications/2%
007/touboul-faugeras:07e}

\bibitem{arnold:83b}
Arnold B (1983) Pareto distributions.
\newblock International Co-operative Pub. House.
\input{arnold:83b}

\bibitem{stoev:06}
Stoev S, Michailidis G, Taqqu M (2006) Estimating heavy-tail exponents through
  max self-similarity.
\newblock Arxiv preprint mathST/0609163 .
\input{stoev:06}

\bibitem{clauset-etal:09}
Clauset A, Shalizi CR, Newman M (2009) Power-law distributions in empirical
  data.
\newblock SIAM Review 51: 661-703.
\input{clauset-etal:09}

\bibitem{barndorff:95}
Barndorff-Nielsen OE, Cox DR (1995) Inference and Asymptotics.
\newblock Chapman and Hall, London.
\input{barndorff:95}

\bibitem{muniruzzaman:57}
Muniruzzaman A (1957) On measures of location and dispersion and tests of
  hypotheses on a pareto population,".
\newblock Bulletin of the Calcuta Statistical Association 7: 115-123.
\input{muniruzzaman:57}

\bibitem{bauke:07}
Bauke H (2007) Parameter estimation for power-law distributions by maximum
  likelihood methods.
\newblock The European Physical Journal B-Condensed Matter and Complex Systems
  58: 167--173.
\input{bauke:07}

\bibitem{clauset-etal:07}
Clauset A, Young M, Gleditsch K (2007) On the frequency of severe terrorist
  events.
\newblock Journal of Conflict Resolution 51.
\input{clauset-etal:07}

\bibitem{stone:74}
Stone M (1977) An asymptotic equivalence of choice of model by cross-validation
  and akaike's criterion.
\newblock Journal of the Royal Statistical Society Series B (Methodological) :
  44--47.
\input{stone:74}

\bibitem{kass-raftery:95}
Kass R, Raftery A (1995) Bayes factors.
\newblock Journal of the American Statistical Association 90.
\input{kass-raftery:95}

\bibitem{grunwald:07}
Gr{\"u}nwald P (2007) The minimum description length principle.
\newblock Cambridge, MA: MIT Press.
\input{grunwald:07}

\bibitem{mood:63}
Mood A, Graybill F, Boes D (1963) Introduction to the Theory of Statistics.
\newblock McGraw-Hill New York.
\input{mood:63}

\bibitem{vuong:89}
Vuong Q (1989) Likelihood ratio tests for model selection and non-nested
  hypotheses.
\newblock Econometrica 57: 307--333.
\input{vuong:89}

\bibitem{neyman-pearson:33}
Neyman J, Pearson E (1933) On the problem of the most efficient tests of
  statistical hypotheses.
\newblock Philosophical Transactions of the Royal Society of London Series A,
  Containing Papers of a Mathematical or Physical Character 231: 289--337.
\input{neyman-pearson:33}

\bibitem{alili-patie-etal:05}
Alili L, Patie P, Pedersen JL (2005) Representations of the first hitting time
  density of an ornstein-uhlenbeck process.
\newblock Stochastic Models 21: 967-980.
\input{alili-patie-etal:05}

\bibitem{schrodinger:15}
Schrodinger E (1915) Zur theorie der fall-und steigversuche an teilchen mit
  brownscher bewegung.
\newblock Physikalische Zeitschrift 16: 289--295.
\input{schrodinger:15}

\bibitem{plesser:99}
Plesser HE (1999) Aspects of signal processing in noisy neurons.
\newblock Ph.D. thesis, Georg-August-Universit{\"a}t.
\input{plesser:99}

\end{thebibliography}
\end{document}